\newcommand{\CLASSINPUTtoptextmargin}{0.75in}
\newcommand{\CLASSINPUTbottomtextmargin}{2cm}
\documentclass[conference,a4paper]{IEEEtran}
\usepackage[utf8]{inputenc}
\usepackage{amsmath}
\usepackage{amsfonts}
\usepackage{amssymb}
\usepackage[english]{babel}
\usepackage{graphicx}
\usepackage{comment}
\usepackage{cite}
\usepackage{cmap}
\usepackage{setspace}
\usepackage{xcolor}
\usepackage{multicol}
\usepackage{float}
\usepackage{tabularx,ragged2e,booktabs}

\newcolumntype{L}{>{\RaggedRight\arraybackslash}X} 

\usepackage{algorithm}
\usepackage{algpseudocode}

\usepackage{graphicx}
\usepackage{float}
\usepackage{wrapfig}

\begin{document}

\title{Cryptoanalysis McEliece-type cryptosystem based on correction of errors and erasures\\
\thanks{The article was prepared within the framework of the Basic Research Program at HSE University.}
}

\author{\IEEEauthorblockN{Kirill Yackushenoks}
\IEEEauthorblockA{\textit{Higher School of Economics} \\
Moscow, Russia \\
kpyakushenoks@edu.hse.ru}
\and
\IEEEauthorblockN{Fedor Ivanov}
\IEEEauthorblockA{\textit{Higher School of Economics} \\
Moscow, Russia \\
fivanov@hse.ru}
}
\maketitle

\begin{abstract}
 Krouk, Tavernier and Kabatiansky proposed new variants of the McEliece
cryptosystem. In this letter, it is shown that cryptosystem based
on correction of errors erasures is
equal to the McEliece cryptosystem with worse parametrs  public key. It will also add an organic extension of the authors' idea, although one that has its flaws...
\end{abstract}


\section{INTRODUCTION} 
McEliece is an excellent cryptosystem with many advantages, but its disadvantages are such as the weight of public keys exceeding thousands of times the weight of other PQ-cryptosystems, prevent it from competing with others.\cite{mceliece1978public} McEliece is an excellent cryptosystem with many advantages, but its disadvantages, such as the weight of public keys exceeding thousands of times the weight of other pq cryptosystems, is a significant disadvantage. This disadvantage and try to remove. The main two ways: 
\begin{enumerate}
    \item Key reduction - use of some constructions from coding theory.
    \item Change the scheme itself or more precisely make the attacker decode not in the sphere, but in the whole set of syndromes.
\end{enumerate}
The authors go the second way. They use a structured error vector. The final error vector consists of both errors and erasures and they also add a codeword.
Such manipulations should make the cryptosystem resistant to attacks with an Information Set Decoding ($ISD$ for short)-like idea.\cite{becker2012decoding} ISD-like attacks force the code length to be taken under the assumption that the desired algorithmic robustness is required $O(2^x) = O(2^{n/20})$. Where n - the length of the code. This estimate is calculated exactly based on the weight of the error yet, but for a rough estimate it is enough, that is, if we want the code persistence $2^{90}$ then we should take $n = 1800, k\approx n/2$. Here we have a matrix 1800 by 900.
They have an interesting idea, but they didn't see through the attack that was cited in the article. In this paper it will be shown that the equation $He^T=s$ has one solution.
Next the article will be in this order in section 2 will be a reminder of how the cryptosystem works, also the attack itself which the authors consider, in section 3 will be a more detailed breakdown and cryptanalysis, in section 4 an idea for improvement is proposed, in 5 implementation, in 6 conclusions.

\section{PRELIMINARY}
\begin{itemize}
\item$G$ - $k \times n$ matrix, which is a generator matrix of a random linear $\lbrack n, k \rbrack$-code $C$ with the minimal distance  $d = d(C)$
\item$M$, $W$ - $n \times n$ nonsingular random matrix
\item$D$ - $n \times n$ diagonal matrix with $r(D)$ ones on its main diagonal, where $r(D)<d$
\item$P_1$, $P_2$ - $n \times n$ permutation matrix
\item$U$ - $n \times k$ matrix of the rank less than 
\item$G_{pub} = GM$ - this generator matrix public code
\item$E_{pub}=(WD(UG + P_1) + P_2)M$ - matrix, which adds errors, erasures and codeword so that an attacker will have trouble decoding it 
\end{itemize}
\begin{equation*}
    y = mG_{pub}+eE_{pub}
\end{equation*}
This equation respond for encryption, where $wt(e) = d/3$.
Now let's talk about decoding or decryption.
We take 
\begin{gather*}
    y=mG_{pub}+eE_{pub}
\end{gather*}
then
\begin{gather*}
    yM^{-1}=mG+e(WD(UG + P_1) + P_2)=\\
    =(m+WDU)G+eWDP_1+eP_2
\end{gather*}
Now, as described above, we need to use decoding to find the codeword of code with generating matrix $(WDU+m)G$. We have
\begin{gather*}
    yM^{-1}-(m+WDU)G=e(WDP_1+P_2)
\end{gather*}
This matrix $(WDP_1+P_2)$ is nonsingular therefore it has $(WDP_1+P_2)^{-1}$. We have vector $e$. A codeword was sent 
\begin{equation*}
    c=y-eE_{pub}
\end{equation*}
From $c$ we obtain $m$.
Firstly, we know place for erasures  now we just need to nullify them and decode without them. Since the error was introduced by weight $d/3$, from the construction of the matrix, it can only become smaller.(If it falls on erasures). So we can decode using the decoding algorithm for $C$ code. We got the codeword sum of two codewords, now if we subtract it from what has been obtained, we get the error vector along with the erasures. Knowing the inverse for the matrix $(WDP_1+P_2)$ that introduces errors and erasures, you can get the vector that the sender contributed.

Now let's talk about the attack. In the paper, the authors consider such an attack:
Eve evaluates an $(n-k)\times n$ matrix $H_{pub}$ such that
\begin{equation*}
    G_{pub}H^{T}_{pub} = 0
\end{equation*}
Evaluates a parity-check matrix for the code $C_{pub}$ with generator matrix $G_{pub}$. It is easy to verify that
\begin{equation*}
   H^T_{pub} = M^{-1}H^T 
\end{equation*}
Where H is some parity-check matrix for the code C. Then Eve evaluates the syndrome $s = yH^T_{pub}$ and tries to solve the following equation
\begin{gather*}  
    mG_{pub}H^T_{pub} + eE_{pub}H^T_{pub} = eH^T_* = s
\end{gather*}
where
\begin{gather*} 
    H^T_*= E_{pub}H^T_{pub} = (WD(UG + P_1) + P_2)MM^{-1}H^T =\\
    =(WDP_1 + P_2)H^T
\end{gather*}
This equation can be considered as syndrome equation for the code $C_*$ with the
parity-check matrix $H_*$ . Note that there is an obstacle for Eve in this way, namely,
the code $C_*$ is not equivalent to the code C as it is for McEliece system. The
minimal distance of the code $C_*$ is unknown and moreover very probably it is
approximately the same as the distance of a random $\lbrack n, k \rbrack$-code what is twice
less than the distance of the initial good code, like Goppa code. The authors write about this in their paper and also note that because of this attack there is no fundamental difference in persistence when using the M matrix, so they will use the permutation matrix.

\section{CRYPTANALYSIS}

The authors choose the number of erasures and errors from the equation, $a \in N$:
\begin{equation*}
    r(D) + 2t + a  = d
\end{equation*}
Now let's look at 2 codes: $C \::\:(H^T, d)$ and $C'\::\: ((WDP_1 ++ P_2)H^T, d')$, where first is the code parity-check matrix and the transpose matrix, and the second is the code minimum distance. It is also worth recalling that the matrix $(WDP_1++P_2)$ - nonsingular. Two things follow from this:
\begin{enumerate}
    \item if $e_1 \neq e_2$ then $e_1WDP_1 \neq e_2WDP_1$
    \item if $e \neq 0$ then $eWDP_1 \neq 0$
\end{enumerate}
Consider the equation below at $e \neq 0$:
\begin{equation*}
    e(WDP_1+P_2)H^T = 0
\end{equation*}
That means $e(WDP_1+P_2)$ is codeword from the $C$ code then $wt(e(WDP_1+P_2))\geq d$.
\begin{gather*}
    wt(e(WDP_1+P_2)) = wt(eWDP_1+eP_2))=\\
    =wt(eWDP_1)+wt(eP_2)=r(D)+wt(eP_2)\geq d\\
    wt(eP_2)\geq r(D) + 2t + a - r(D)\\
    wt(eP_2)\geq 2t + a\\
    wt(e)\geq 2t + a
\end{gather*}

This is the worst case, there is a situation where erasures and errors do not intersect, otherwise there will be even more $d$. That is, this is the lower bound. From what has been written above, it follows that $d' \geq 2t + a$ and $a \geq 1$ so the code $C'$ can decode $wt(e) \leq t$ clearly. It is worth noting again that in this cryptosystem $ t = d/3$ and in McEliece $t = d/2$, $ISD$ is less difficult, there are larger and heavier matrices to store in this cryptosystem.

That's the second place to launch an attack from if the matrix M is changed to a permutation one(maybe even the first ):
\begin{equation*}
    E_{pub}=((WDUG + WDP_1) + P_2)P
\end{equation*}
If you look at the first summand, you realise that it is a subcode, also that it can have at most $2^{r(D)}$ codewords. For the example from the article, it's $2^{33}$.
\begin{equation*}
    e_1E_{pub}+e_2E_{pub}=(e_1+e_2)(WDUG+WDP_1 + P_2))P
\end{equation*}
Our goal is to find 2 errors such that the code words involved in creating the error are the same, then 
\begin{equation*}
    (e_1+e_2)(WDP_1P + P_2P)=(e_1+e_2)WDP_1P + (e_1+e_2)P_2P
\end{equation*}
or we need to find one vector: 
$eWD=0$ then 
\begin{equation*}
    eE_{pub}=eP_2P, \text{    }  P^*=P_2P,
\end{equation*}
 it should also be noted $wt(eP_2P)=t$. 
 
Task: restore $P^*$. Algorithm 1 will help with this.
\begin{algorithm}
\caption{Algorithm to restore $P^*$}\label{alg:cap}
\begin{algorithmic}
\Require $t$, $E_{pub}$
\Ensure $P^*$
\State $P^* \gets [z]^n$ 
\State $count \gets 0$
\While {$count < 2n$} 
    \State $e \gets$ generate random $wt(e) = t$
    \State $b \gets eE_{pub}$
    \State $l \gets supp(b)$
    \If{$wt(b) = t$}
        \State $count \gets count + 1$ 
        \For{$j \gets 1 \text{ to } n$} \Comment{including $n$}
            \If {$e[i] = 1$} 
                \If {$quantity(P^*[j]) = 1$ and $P^*[j] = z$}
                    \State $P^*[j] = l$               
                \ElsIf {$quantity(P^*[j]) > 1$}
                    \State $P^*[j] = set(P^*[j])$ \& $set(l)$ 
                \EndIf
            \EndIf            
        \EndFor
    \EndIf
\EndWhile
\end{algorithmic}
\end{algorithm}

This algorithm is here to show schematically the process. It can be improved, but further there will be ISD attack and its complexity is much higher, so you can not bother much. $2n2^{r(D)}=2^{r(D)+1+log_2(n)}$ - This is how many times the while loop will run. For parameters are $2^{44}$. For loop will run 2n times. We know that $E_{pub} = (WD(UG + P_1) + P_2)P$ and $P_2P$ next 
 \begin{gather*}
     E_{pub}=WD(UG + P_1)P + P_2P\\
     E_1=WD(UG + P_1)P\\
     E_2 = P_2P
 \end{gather*}
Now we need to develop an attack using the knowledge of $E_2$.
The channel transmits $y = mG_{pub} + eE_{pub}$. Here's what we do 
\begin{gather*}
    yH_{pub}^T=(WDP_1 + P_2)H^T,\\
    where\\
    H_{pub}^T = P^{-1}H^T.
\end{gather*}
There are erasures and errors in the same matrix, but if you use the knowledge about $E_2$ then 
\begin{gather*}
    A_1 = E_1H_{pub}^T=WDP_1H^T\\
    A_2 = E_2H_{pub}^T=P_2H^T.
\end{gather*}
\begin{equation*}
    yH_{pub}^T=e(A_1+A_2)=eA_1+eA_2= 
\end{equation*}
\[ = \begin{bmatrix}
         e & e  
     \end{bmatrix}
     \times
     \begin{bmatrix}
         A_1\\
         A_2  
     \end{bmatrix}
      = yH_{pub}=s=(s_1||s_2) \]

It is also worth noting that $rank(A_1) = r(D)$. The matrix $\lbrack A_1 || A_2 \rbrack$ can be reduced to the form shown in Figure 1(quasi-systematic view).

\begin{figure}[h]
\centering
\includegraphics[width=0.8\linewidth]{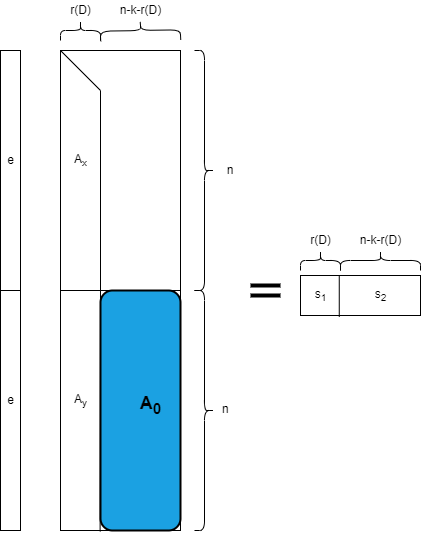}
\caption{Quasi-systematic view}
\label{fig:mpr}
\end{figure}

Figure 2 shows a schematic view of the $ISD$ for the matrix $A_0$.

\begin{figure}[h]
\centering
\includegraphics[width=0.8\linewidth]{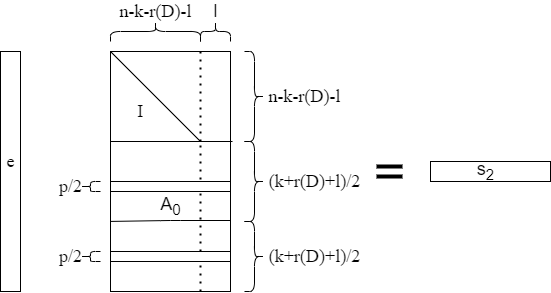}
\caption{Schematic view for $ISD$ attack}
\label{fig:mpr}
\end{figure}

Now if you look at the matrix, then $eA_0=s_2$, for this matrix we need to run the $ISD$-algorithm. The input is $A_0$, $s_2$, $t$, the algorithm will search for $t$ and less weight.
Probability compared to the original McEliece:
\begin{equation*}
    \frac{P(\text{This cryptosystem})}{P(\text{McEliece}))}= \frac{ \binom{n-k-l-r(D)}{d/3-p} \binom{(k+l+r(D))/2}{p/2}^2 \binom{n}{d/2}}{\binom{n-k-l}{d/2-p} \binom{(k+l)/2}{p/2}^2 \binom{n}{d/3}} 
\end{equation*}
If we substitute the numbers for McEliece $[1024,524]$, $t = 50$ and compare with $[1023,490]$, $t=r(D)=33$ for this cryptosystem  $\approx2^{18}$.
Thus with probability more than 60\% we can find vector e, if not we repeat, then the probability will be 84\%, if not, so on and so forth. vector e is singular otherwise a legitimate user cannot decrypt the message too, since such a matrix construction has separated erasures from errors. You can check, of course, you have to $eA_x+eA_y=s_1$.
The error weight t was found in this way.

\section{REMEDIAL IDEAS}
Increasing $r(D)$ helps to solve these 2 problems. But in the usual case $t$ and $r(D)$ binds the $2t+r(D)<d$. Therefore, it is proposed to build the matrix $G = G_0 || G_1$ where $G_0$ - $k \times n$ code matrix with an efficient decoding algorithm that can fix $wt(e)=\frac{d-1}{2}$,  $G_1$ - $k \times v$ the matrix of another code. The idea of putting erasures on $G_1$,  $r(D)= v$.
\begin{itemize}
    \item Weakness: is $R$ more than before $R = \frac{k}{n+v}$.
    \item Positive: in a regular code scheme, the following can be said about the number of corrected errors depending on n $t \approx \frac{0.1n}{2}=0.05n$, but with the addition of code for erasures $t=0.5n$. Here by error he means how many of them will be at random $e$ introduced into the code by erasures.
\end{itemize}
We also suggest another way to close from the second attack, a more graceful one. 
\begin{equation*}
    E_{pub} = (WUG + WDP_1 + P_2)P
\end{equation*}
Then there will be no second attack provided that $WDP_1 + + P_2$ has an inverse matrix and $e_{pub}$ - singular.

\section{Implementation}
We ran a trial implementation
of the modernised this cryptosystem and the attack algorithm in
Section III using the Python language.
It should be noted that the
decryption does not use decoding of the C code and
therefore we can treat C as a random code rather than a
well-designed code with efficient decoding such as Goppa or BCH code.
Source code: {\it https://github.com/Persequentes/
search\_matrix\_permutation\_on\_E\_pub}

\section{Conclusion}
We would like to point out that the authors have an interesting idea, but it has been shown that a prototype cryptosystem
based on error and erasure correction is equivalent to the McEliece cryptosystem, but with a lower complexity factor of the exponent. There is also one idea that helps to close from the second attack, but it is harder to close from the first with this idea, you need to choose the right x. If r(D) is large enough, then you can try to switch to probabilistic decoding and increase the error vector.

\section{Acknowledgement}
The article was prepared within the framework of the Basic Research Program at HSE University in 2023.

\bibliographystyle{IEEEtran}
\bibliography{IEEEabrv,IEEEexample}
\end{document}